\documentclass{iaus}
\usepackage{graphicx}

\title[Early Type Galaxies in the Mid Infrared] 
{ Early Type Galaxies in the Mid Infrared:\\
a new flavor to their stellar populations  
\thanks{This work is based on observations made with the Spitzer Space Telescope, which
is operated by the JPL, Caltech
under a contract with NASA. }}

\author[Bressan, Panuzzo, Vega et al.]   
{A. Bressan$^{1,2,3}$,
 P. Panuzzo$^1$,
 O. Vega$^3$,
 L. Buson$^1$, 
 M. Clemens$^1$, 
 G.L. Granato$^{1,2}$, 
 R. Rampazzo$^1$,
 L. Silva$^4$, 
 J.R. Valdes$^3$
  }
\affiliation{$^1$INAF Osservatorio Astronomico di Padova, vicolo dell'Osservatorio 5, 35122 Padova, Italy
\break email: alessandro.bressan@oapd.inaf.it\\[\affilskip]
$^2$SISSA, via Beirut 4, 34014, Trieste, Italy\\[\affilskip]
$^3$ INAOE, Luis Enrique Erro 1, 72840, Tonantzintla, Puebla, Mexico \\[\affilskip]
$^4$ INAF Osservatorio Astronomico di Trieste, Via Tiepolo 11, I-34131 Trieste, Italy}

\pubyear{2007}
\volume{241}  
\pagerange{1--4}
\date{?? and in revised form ??}
\jname{Stellar Populations as Building Blocks of Galaxies}
\editors{A. Vazdekis and R. Peletier}
\begin{document}

\maketitle

\begin{abstract}
The mid infrared emission of early type galaxies traces
the presence of intermediate age stellar populations as well as even
tiny amounts of ongoing star formation.
Here we discuss high S/N {\it Spitzer} IRS spectra of a sample of Virgo 
early type galaxies, with particular reference to NGC 4435.  
We show that, by combining  mid
infrared spectroscopic observations with existing broad band fluxes,
it is possible to obtain a very clean picture of the nuclear activity in
this galaxy.
\end{abstract}

\firstsection 
\section{Introduction}
With the advent of the {\it Spitzer Space Telescope} new frontiers have been opened
in the study of stellar population content of early-type galaxies (ETGs)
and, in particular,  the ability to quantify the occurrence and strength
of the {\it rejuvenation} episodes.
By means of mid infrared (MIR) observations it is possible to 
detect the presence of intermediate age stellar populations 
in passively evolving galaxies, and measure 
even tiny amounts of ongoing star formation activity.

Bressan, Granato \& Silva (1998) suggested that the MIR spectral region of old
and intermediate age stellar populations should be affected by the presence of
mass-losing oxigen-rich AGB giants. Their integrated emission around 10$\mu$m
should be clearly seen in passively evolving galaxies; its analysis, in
combination with UV, optical and NIR observations, should provide 
accurate age-metallicity ranking,  unbiased by the age-metallicity
degeneracy.

Moreover, ongoing star formation can be  easily
detected in the MIR, from the presence of prominent
emission features such as PAHs and atomic or molecular 
emission lines (e.g. Kaneda et al. 2005, Bressan et al. 2006a,b,
Panuzzo et al. 2007).
Last but not least, MIR nebular lines constitute a strong diagnostic 
to disentangle star formation and AGN activity and 
they also allow a {\sl direct and perhaps unique} determination of
the chemical abundance of the surrounding gas (Panuzzo et al. 2007).

For the above reasons we begun a systematic study of the properties of 
ETGs in the mid infrared spectral region with the {\it Spitzer Space Telescope}.
Here we report on the results obtained with {\it Spitzer} IRS
(Houck et al. 2004) 
MIR spectroscopic observations of a sample of ETGs
in the Virgo cluster (Bressan et al 2006a,b; Bressan et al 2007).



\section{Early-type galaxies in the mid infrared.}
Eighteen ETGs among those that define the colour-magnitude relation of the 
Virgo cluster (Bower, Lucy \& Ellis 1992) were observed in standard staring mode 
with the low resolution IRS modules between 5 and 20$\mu$m, in January and July 2005.
The calibration and spectra extraction procedures are discussed in detail in Bressan et al. (2006a). 
The spectra of these galaxies are shown in Bressan et al. (2006a) and 
Bressan et al. (2007).

For thirteen galaxies (76\%) of our sample, the MIR spectrum is
characterized by the presence of a broad emission features
above 10$\mu$m, {\sl without any other narrow emission feature}.
The analysis of the IRS spectra indicates that the {\sl 10$\mu$m feature}
has an extended spatial distribution; moreover its spatial distribution is
consistent with that obtained below 8$\mu$m, where the spectra are
dominated by stellar photospheres.
This result has been confirmed by the analysis of {\it Spitzer} IRS Peak-Up imaging observations in the blue (16$\mu$m) filter of selected galaxies (Annibali et al. in preparation). It is also in agreement with previous ISOCAM observations  
that indicated spatially resolved emission at both 6.7 and 15 $\mu$m
(Athey et al. 2002, Ferrari et al 2002, Xilouris et al. 2004).
In view of these considerations and based on 
preliminary fits with our models of passively evolving old simple stellar populations,
we argued that we have detected the 10$\mu$m features, due to silicate emission from 
the circumstellar envelopes of mass losing AGB stars, as
predicted by Bressan et al. (1998). 
Bressan et al. (2007) have recently shown that
the 10$\mu$m feature observed  in early type galaxies
is similar in shape but about a factor four
larger than the {\it semi empirical} one obtained for the globular cluster 
47 Tuc, consistent with a metallicity variation of the same order. 
We are now computing new isochrones and SSP models
that account for a more realistic description of the AGB phase and of their
dusty envelopes.
\begin{figure}
\centerline{ \includegraphics[width=0.5\textwidth]{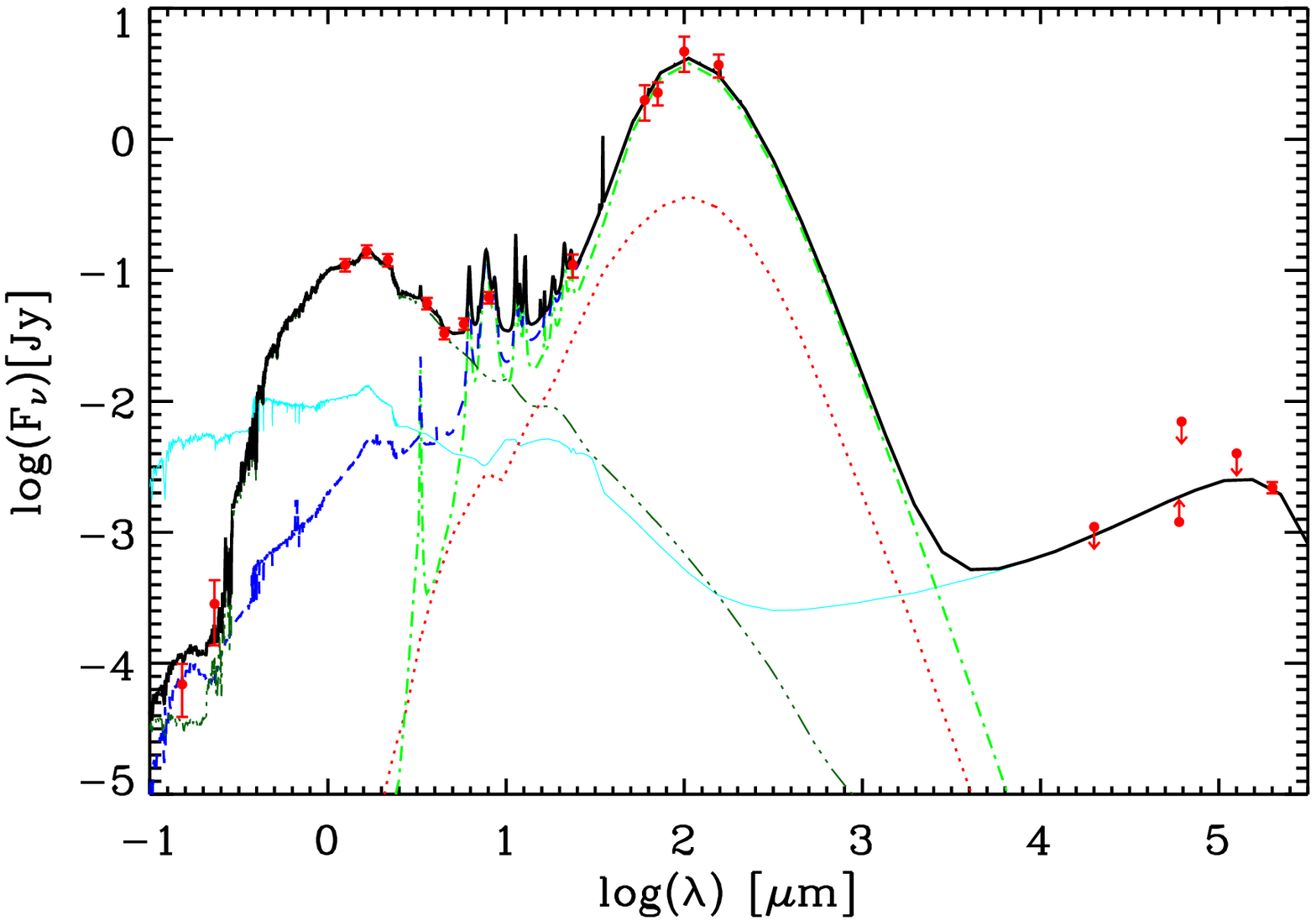}
 \includegraphics[width=0.5\textwidth]{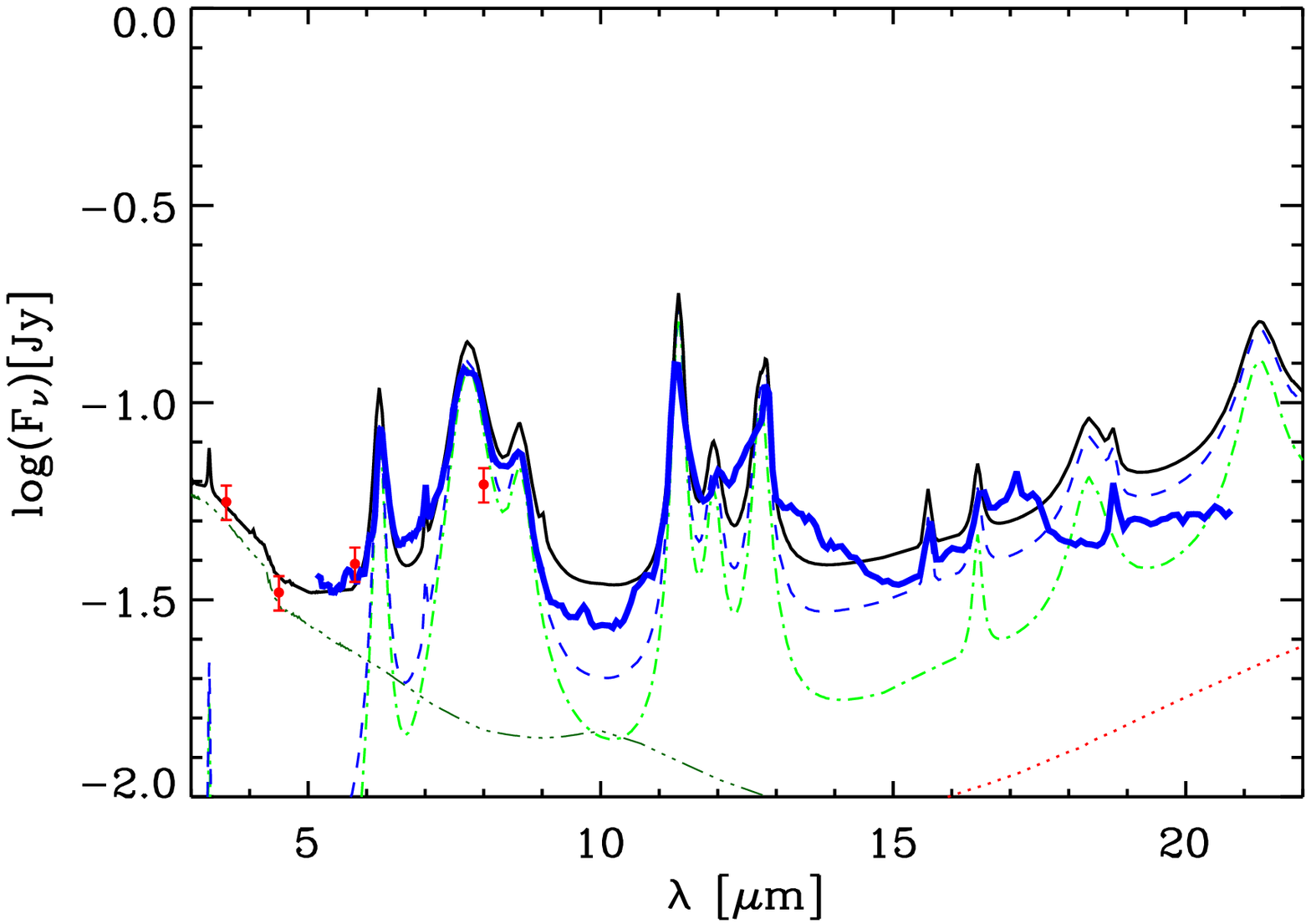}}
\caption{Comparison between the observed SED of the central region of
NGC~4435 and the GRASIL model. 
The thick solid line
represents the model for the total SED, i.e. the starburst
component plus the old stellar component; the three dots-dashed dark green
line represents the contribution from the old stellar population, and
the dashed blue line represents the total contribution from the burst
of star formation, the dotted red line represents the emission from
molecular clouds, the dot-dashed green line represents the diffuse medium
emission and the thin solid cyan line denotes the emission from
stars of the starburst component without applying the extinction from dust. The
filled red circles are the broad band data. 
\textit{Left}: Comparison from 0.1$\mu$m to 100 MHz.
\textit{Right}: 
Comparison for the MIR wavelengths.
The thickest solid blue line represents the IRS \emph{Spitzer} spectrum.
} 
\label{active}
\end{figure}

Among bright Virgo cluster ETGs observed by our team, four galaxies (24\%)
show various levels of activity.
NGC 4636 (optically classified as a LINER) shows low ionization
emission lines ([ArII]7$\mu$m, [NeII]12.8$\mu$m,  [NeIII]15.5$\mu$m and [SIII]18.7$\mu$m)
on a continuum similar to other passive galaxies.
NGC 4486 (M87) shows the same emission lines on a continuum dominated by the 
AGN emission above 8$\mu$m. The broad continuum feature above 10$\mu$m 
in M87 could be caused by silicate emission from the dusty torus 
(Siebenmorgen et al. 2005, Hao et al. 2005).
NGC 4550 shows some PAH emission features while the MIR SED of NGC 4435 
is characteristic of 
a star forming object.


\section{The panchromatic SED of NGC 4435}
NGC 4435 is an S0 galaxy interacting with NGC 4438 and it hosts a
circumnuclear disk.
Panuzzo et al. (2007) combined
the \emph{Spitzer} IRS spectra of NGC 4435 with IRAC and MIPS 
archival data and existing broad band
measurements from X-ray to radio wavelengths to obtain an accurate
panchromatic spectral energy distribution (SED) of this galaxy.
The SED was analysed with GRASIL (Silva et al. 1998) and
well reproduced at all wavelengths. The analysis
shows that the circumnuclear disk experienced a
burst of star formation activity which is now fading.

The IRS data themselves provide precise answers on 
important questions such as the nature of the nuclear activity suspected 
from optical (Ho et al. 1997) and X-ray (Machacek et al. 2004) observations, 
and the metallicity of the gas in the circumnuclear disk.
We fail to detect any high excitation nebular emission lines in the IRS spectrum;
the [NeIII]15.5/[NeII]12.8 ratio constrains the contribution 
of a possible AGN to the ionizing flux to be less than 2\%.
The upper limit on the temperature derived from H$_2$ S(1) and S(2)
rotational lines is lower than expected for AGN excitation and
PAH features are well reproduced by star formation models.
Moreover, the X-ray emission is within the range expected from
X-ray binaries in an advanced phase of the starburst.

As for the metallicity of the nuclear disk,
the comparison of observed MIR nebular lines with those predicted by 
the GRASIL model (Panuzzo et al 2003) indicates that 
it is almost solar. This is one of the first accurate 
{\sl direct} estimates of the gas metallicity in ETGs.

The age of the starburst, $\sim$180 Myr, corresponds to the
epoch of the onset of the interaction with NGC~4438 derived from dynamical simulations
(Combes et al. 1988).
The mass of stars born during the starburst ($\sim 1.22\times10^8~M_\odot$)
amounts to about 1.5\% of the stellar mass sampled by the central 
5 arcsec aperture.

\section{Conclusions}
We have obtained with \emph{Spitzer} IRS mid infrared spectra of ETGs
selected along the colour-magnitude relation of the Virgo cluster.

The mid infrared SED of most of our ETGs shows a clear broad emission
around 10$\mu$m and longward as predicted in Bressan et al. (1998)
which is likely due to dusty mass losing  AGB stars.
In the remaining fraction of galaxies (24\%) we detect signatures of activity
at various levels.
The analysis of the IRS spectrum of NGC 4435
testifies to the superb capability of 
Spitzer to probe the nature of this type of activity and
supports the notion that ETGs with 
relatively strong  hydrogen absorption features 
are due to recent small rejuvenation episodes, rather than
being the result of delayed galaxy formation (Bressan et al. 1996).

\begin{acknowledgments}
A. B., G.L. G. and L. S.  thank INAOE for warm hospitality.
\end{acknowledgments}

\begin{discussion}

\discuss{Meixner}{I'm happy to see that someone is including dust in their models.
You say that the 10$\mu$m emission in the elliptical is extended. Is your
interpretation that the AGB population is extended? Is the extended emission
consistent with the AGB population as traced by near-IR photometry?}

\discuss{Bressan}{Our IRS data analysis pipeline allows an accurate check
of the spatial distribution of the emission. In the passive galaxies,
the emission looks extended at all wavelengths.
The spatial extension is consistent with
IRAC and 2MASS images. The preliminary analysis of our {\it Spitzer} IRS Peak-Up imaging observations confirms the extension also at 16$\mu$m.
This indicates that the 10$\mu$m excess is of stellar origin, likely from the 
extended population of AGB stars.}

\discuss{Renzini}{One may expect that dust particled from AGB stars are destroyed by sputtering due to interaction with the hot, X-Ray emitting ISM of these galaxies. Is there
any correlation of the strength of the emission with the X-ray luminosity
of these galaxies?}

\discuss{Bressan}{We did not compare with the X-ray luminosity yet. However
the dust distribution from stationary circumstellar envelopes is proportional 
to r$^{-2}$ and the emission is dominated by the innermost layers,
where dust is likely not affected by interaction with the hot ISM.} 
\end{discussion}

\end{document}